\documentclass[letter,preprint]{jpsj2}
\usepackage[]{graphicx,color}
\usepackage{bm}

\def\runauthor{Yoshihiro Shimomura}

\title{%
Non-Degenerate Ground State \\
in the Antiferromagnetic Double-Exchange Model \\
on a Triangular Lattice. 
}

\author{%
Yoshihiro {\sc Shimomura}\thanks{E-mail address: shimomura@phys.aoyama.ac.jp},
Shin {\sc Miyahara} and Nobuo {\sc Furukawa} 
}

\inst{%
Department of Physics and Mathematics, Aoyama Gakuin University, 
Sagamihara 229-8558, Japan
}

\recdate{\today}

\abst{%

In order to study effects of frustration in an itinerant electron system,
we investigate ground states  of the antiferromagnetic
double-exchange model on a triangular lattice.
In this model, pseudo-spins are coupled to electron transfer integrals
in such a way that antiparallel configurations of pseudo-spins
gain kinetic energies.
Although the antiferromagnetic Ising model on a triangular
lattice shows  macroscopic dengenerate ground states,
the present model shows that
the degeneracy is
lifted due to long-range natures of the 
 double exchange interactions.
Spin ordering at the ground state is also discussed.

}

\kword{%
antiferromagnetic double-exchange model, triangular lattice, 
geometrical frustration, degeneracy
}

\begin{document}
\maketitle

One of the interests in the study of frustrated systems
is to investigate anomalous ground states where
conventional long-range orders are suppressed.
In some systems,  disordered ground states with macroscopic degeneracies 
are observed.
Since such states are thermodynamically
unstable, singular responses to perturbations may be observed.
Typical examples are antiferromagnetic (AF) Ising models 
on  geometrically frustrated lattices, e.g., on 
 triangular\cite{Wannier_50}, 
 kagom\'e\cite{Syozi_51,Kano_53} and 
 pyrochlore\cite{Anderson_56} lattices.

Another example which shows the macroscopically degeneracy is the
spin ice systems on kagom\'e and pyrochlore lattices.
In these systems, spins have uniaxial
anisotropies so that they take either of  ``inward'' and ``outward''
states. We define the pseudo-spin 
$\tau_i$ by $\tau_i = 1$ $(-1)$ if the spin on the $i$-th site points outward
(inward).
In spin ice systems, 
ferromagnetic interaction between spins can be mapped to AF 
interaction between pseudo-spins.
Therefore, the spin ice model 
with ferromagnetic nearest neighbor (n.n.) Ising  interactions 
exhibits  macroscopically degenerate ground states.\cite{Bramwell_01_science}

When such  macroscopic degeneracies are lifted
by residual interactions, peculiar states such as
cluster ordered states emerge as the ground states.
For instance, 
when dipolar long-range interactions are introduced
in the spin ice model on a pyrochlore lattice,
a spin cluster ordered state is observed at 
sufficiently low temperatures.\cite{Melko_01}

Recently, the authors have investigated  frustrated itinerant electron
systems. 
One of the motivation to investigate
such systems 
is to study anomalous ground states
which might exhibit unconventional transport and optical properties.
Another interesting point is to understand 
how the nature of interactions in itinerant electron systems are
different from those in localized spin systems.
We have introduced  a double exchange spin ice (DESI) model 
in the strong Hund's coupling limit on a kagom\'e lattice, 
and observed a spin cluster ordered state which is 
named the dodecamer ordered phase.\cite{Shimo_04}
In the DESI system, 
interactions are ferromagnetic due to the double exchange
mechanism,\cite{Zener_51,Anderson-Hasegawa_55}
and uniaxial anisotropies are introduced for local spins.
Due to the spin ice mechanism, 
there exist AF interactions between 
the pseudo-spins defined on the kagom\'e lattice.
The DESI model 
exhibits a translational symmetry broken state,
although the analogous spin ice system has disordered
ground state with the macroscopic degeneracy.
Thus, differences in the nature of interactions
make  qualitatively different results in the
behaviors of the ground states.

The realization of the spin ice mechanism 
enables us to construct a double exchange system 
with the AF interaction between pseudo-spins. 
It is difficult to introduce such mechaisms 
on geometrically frustrated lattices 
other than the kagom\'e and the pyrochlore lattice. 
However, in order to investigate the effects of 
the kinetics of electrons and the frustration, 
let us simply extend the model to general cases 
on various frustrated lattices. 
We define an antiferromagnetic double-exchange (AF-DE) model 
in the form, 
\begin{align}
\hat{H} = -\sum_{\langle i,j \rangle} t(\tau_{i}, \tau_{j})
\left( 
c_{i}^{\dagger} c_{j} + h.c.
\right), \label{eq:AFDEmodel}
\end{align}
where
the summation is taken over all pairs of the n.n. sites, and
\begin{align}
t(\tau_i, \tau_j) = 
\begin{cases}
t \qquad (\tau_i = - \tau_j), \\
t^{\prime} \qquad (\tau_i = \tau_j).
\label{eq:teff}
\end{cases}
\end{align}
Here, $t$ and $t^{\prime}$ are strong and weak transfer integrals, 
respectively, which satisfies $0 \le |t'/t| \le 1$.
It is easily found that the AF-DE model 
on a geometrically frustrated lattice is a frustrated electron system.

The DESI model corresponds to the AF-DE model on a
kagom\'e lattice with $t'/t = 1/\sqrt{3}$.\cite{Shimo_04}
For a given spin configuration, the energy 
 of the AF Ising model is determined by
the  number of AF bonds, $N_{\rm AF}$.
Similarly, in the AF-DE model, $N_{\rm AF}$ gives the
total number of strong-hopping bonds.
Increase of $N_{\rm AF}$
roughly gives the kinetic energy gain.
Therefore, there exist AF interactions between pseudo-spins
in the AF-DE model.
To be precise, however,
pseudo-spin interactions in 
the AF-DE model are not identical to those in the AF Ising model.
The energy of the AF-DE model for a given pseudo-spin configuration
$\{\tau_i\}$ is given by
\begin{align}
E_{\rm K}(\{\tau_i\}) = -\sum_{\langle i,j \rangle} t(\tau_{i}, \tau_{j})
\left\langle
c_{i}^{\dagger} c_{j} + h.c.
\right\rangle_0, \label{eq:AFDE-EG}
\end{align}
where $\langle\cdots\rangle_0$ represents 
an expectation value at the ground state.
If  $\langle c_{i}^{\dagger} c_{j} + h.c.\rangle_0= {\rm const.}$,
$E_{\rm K}$ depends only on $N_{\rm AF}$
so that the AF-DE model can be mapped to the AF Ising model exactly.
However, this is not the case, since 
the expectation value of electron hopping
 $\langle c_{i}^{\dagger} c_{j} + h.c.\rangle_0$ 
depends on both eigenenergies and wavefunctions of the Hamiltonian.
Due to the extended nature of the wavefunctions of the itinerant 
electrons,  $\langle c_{i}^{\dagger} c_{j} + h.c.\rangle_0$ cannot
 be expressed by local pseudo-spin configurations in a simple manner.
Namely, a change in a pseudo-spin on the $k$-th site $\tau_k$ affects
the wavefunctions and hence $\langle c_{i}^{\dagger} c_{j} + h.c.\rangle_0$
in neighboring sites.
Therefore, $E_{\rm K}$ depends on the global lattice structure
which consists of strong and weak bonds.
As a result, behaviors of the AF-DE model at low temperatures 
might be different from those of the AF Ising model.

In this paper, 
in order to investigate non-trivial ground states 
in the frustrated itinerant electron systems,
the AF-DE model on the triangular lattice is studied. 
For simplicity, we take $t'=0$ throughout this paper.
We also set the chemical potential $\mu = 0$. 
Results for $t'$, $\mu \ne0$ will be reported elsewhere.

\begin{figure}[ht]
\begin{center}
\includegraphics[width=6cm,keepaspectratio]{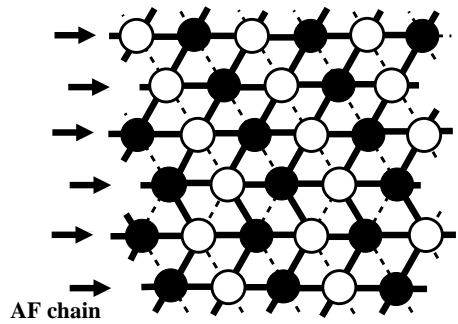}
\end{center}
\caption{
An AF chain stacked state of the AF Ising model, 
where each AF chain is indicated by arrows. 
Open and closed circles represent the up- and the down-spins, respectively. 
Solid lines represent bonds with AF configuration, 
while dashed lines represent bonds with ferromagnetic configuration. 
}
\label{fig:GSAFIsing}
\end{figure}

\begin{figure}[ht]
\begin{center}
\includegraphics[width=8cm,keepaspectratio]{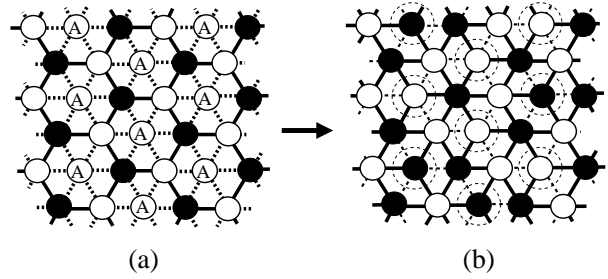}
\end{center}
\caption{
Three sublattice structure in the AF Ising model. 
Open and closed circles represent the up- and the down-spins, respectively. 
(a) Lattice sites consist of 
three sublattices, ``up-spin'', ``down-spin'' and ``$A$'' sublattices. 
Directions of spins on $A$ sublattice sites are arbitrary. 
(b) An example of the three sublattice structures. 
Solid lines represent bonds with AF configuration, 
while dashed lines represent bonds with ferromagnetic configuration. 
$A$ sublattice sites are indicated by dashed circles. 
$N_{\rm AF}$ does not depend on the spin states on the $A$ sublattice 
since such spins always have 
three up-spins and three down-spins at n.n. six sites. 
}
\label{fig:GSAFIsing_ent}
\end{figure}

The ground states in the AF Ising model which maximize $N_{\rm AF}$ 
may be considered as good candidates for those in the AF-DE model. 
Among all ground states in the AF Ising model, 
two types of them are shown 
in Fig. \ref{fig:GSAFIsing} and Fig. \ref{fig:GSAFIsing_ent}. 
One is a state constructed by stacking the AF chains 
(AF chain stacked state) 
[Fig. \ref{fig:GSAFIsing}], 
and the other is one which has the three sublattice structure 
[Fig. \ref{fig:GSAFIsing_ent}(a)] 
, i.e., 
the up-spin sublattice, the down-spin sublattice and the $A$ sublattice. 
As shown in Fig. \ref{fig:GSAFIsing_ent}(b), 
each spin state on the $A$ sublattice 
can be determined independently since $N_{\rm AF}$ does not depend on 
spin states on $A$ sublattice. 
Hereafter, we treat the AF-chain stacked states 
and the three sublattice states 
as the possible candidates for the ground state in the AF-DE model 
and calculate energies of these states for small systems.

In order to check whether the degeneracy in the Ising system 
can be lifted by the kinetics of electrons in the AF-DE model or not, 
we have calculated energies of the AF chain stacked states and 
the three sublattice states for the finite systems. 
In order to extrapolate the energy in the thermodynamic limit 
from that in the finite systems, 
twisted boundary conditions have been applied, and 
the energy of the system has been calculated by averaging over phases. 
This procedure extrapolates the energy to that for a system 
in the thermodynamic limit with a finite-sized supercell. 
The density of states (DOS) in each spin configuration 
has also been calculated by this method.

\begin{figure}
\begin{center}
\includegraphics[width=8cm,keepaspectratio]{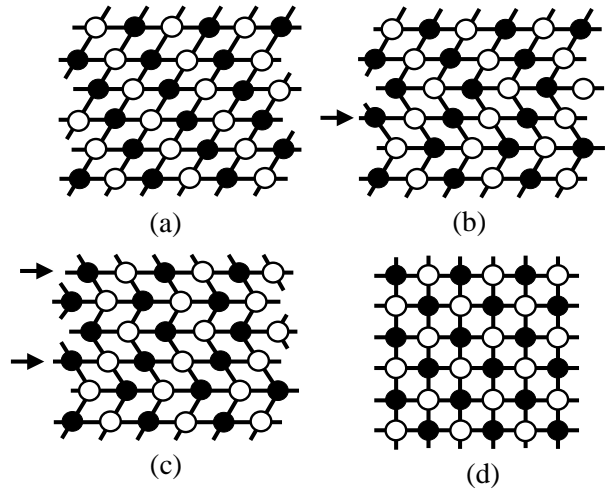}
\end{center}
\caption{
AF chain stacked states in the AF-DE model.  
Open and closed circles represent the up- and the down-spins, respectively. 
Solid lines represent bonds created by the electron hopping. 
The patterns (b) and (c) are obtained from the pattern (a) 
by flipping the AF chain indicated by the arrows. 
In terms of the electron hopping, 
these patterns are topologically identical to the regular square lattice (d), 
although spin configurations are different from each other. 
}
\label{fig:bondsquare}
\end{figure}

First, let us discuss the AF chain stacked states in the AF-DE model. 
Considering the electron hopping, it is found that 
the degeneracy in the AF chain stacked state cannot be lifted as follows. 
Since electrons can only hop between the up- and the down-spins, 
the AF chain stacked states can effectively be mapped to 
an electron system on a lattice which is created by disconnecting 
ferromagnetic bonds. 
Several AF chain stacked states in the AF-DE model 
are shown in Fig. \ref{fig:bondsquare}. 
For example, the spin configuration in Fig. \ref{fig:bondsquare}(b) 
is obtained from that in Fig. \ref{fig:bondsquare}(a) 
by flipping the AF chain indicated by the arrow. 
Other configurations [e.g., Fig. \ref{fig:bondsquare}(c)] 
can be obtained in the same manners. 
Although spin configurations are different from each other, 
these AF chain stacked states are topologically equivalent 
to the square lattice as long as the electron hoppings are concerned 
[see Fig. \ref{fig:bondsquare}(d)]. 
In this way, in the thermodynamic limit, 
all AF chain stacked states are regarded as the identical square lattice, 
and thus, have the same energy.

\begin{figure}
\begin{center}
\includegraphics[width=7cm,keepaspectratio]{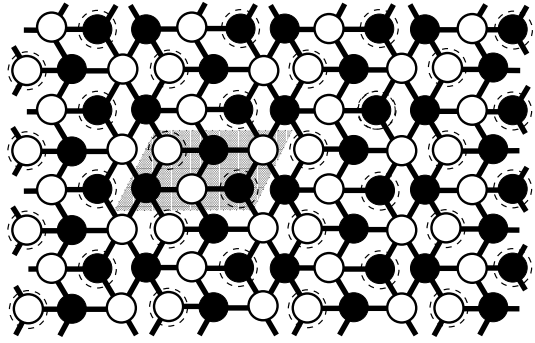}
\end{center}
\caption{The lowest energy state in the three sublattice with 
the $6 \times 6$ supercell in the AF-DE model. 
The dashed circles represent sites on the $A$ sublattice. 
The shadowed region is a unit cell including six sites.
}
\label{fig:GSAFDE}
\end{figure}

Next, we consider the three sublattice states. 
The energy of all possible spin configurations 
within the three sublattice states which consist of 
supercells with $6 \times 6$ sites in the thermodynamic limit 
($6 \times 6$ supercells) has been calculated. 
Note that there exist $2^{12}$ configurations for $A$ sublattice spins 
in this case. 
As a result, 
we have found that the degeneracy is lifted by the kinetics of electrons. 
The smallest energy state with the $6 \times 6$ supercell 
in the thermodynamic limit is shown in Fig. \ref{fig:GSAFDE}. 
Such a periodic state is determined uniquely 
except for the trivial degeneracy.  
Concerning electron hopping, 
each spin configuration in the three sublattice states 
can be mapped to a certain lattice different from each other. 
It is natural that the degeneracy 
in the three sublattice state is lifted by the electron motions, 
since the kinetic energy of electrons is strongly dependent to 
the path of electrons.

\begin{figure}
\begin{center}
\includegraphics[width=8cm,keepaspectratio]{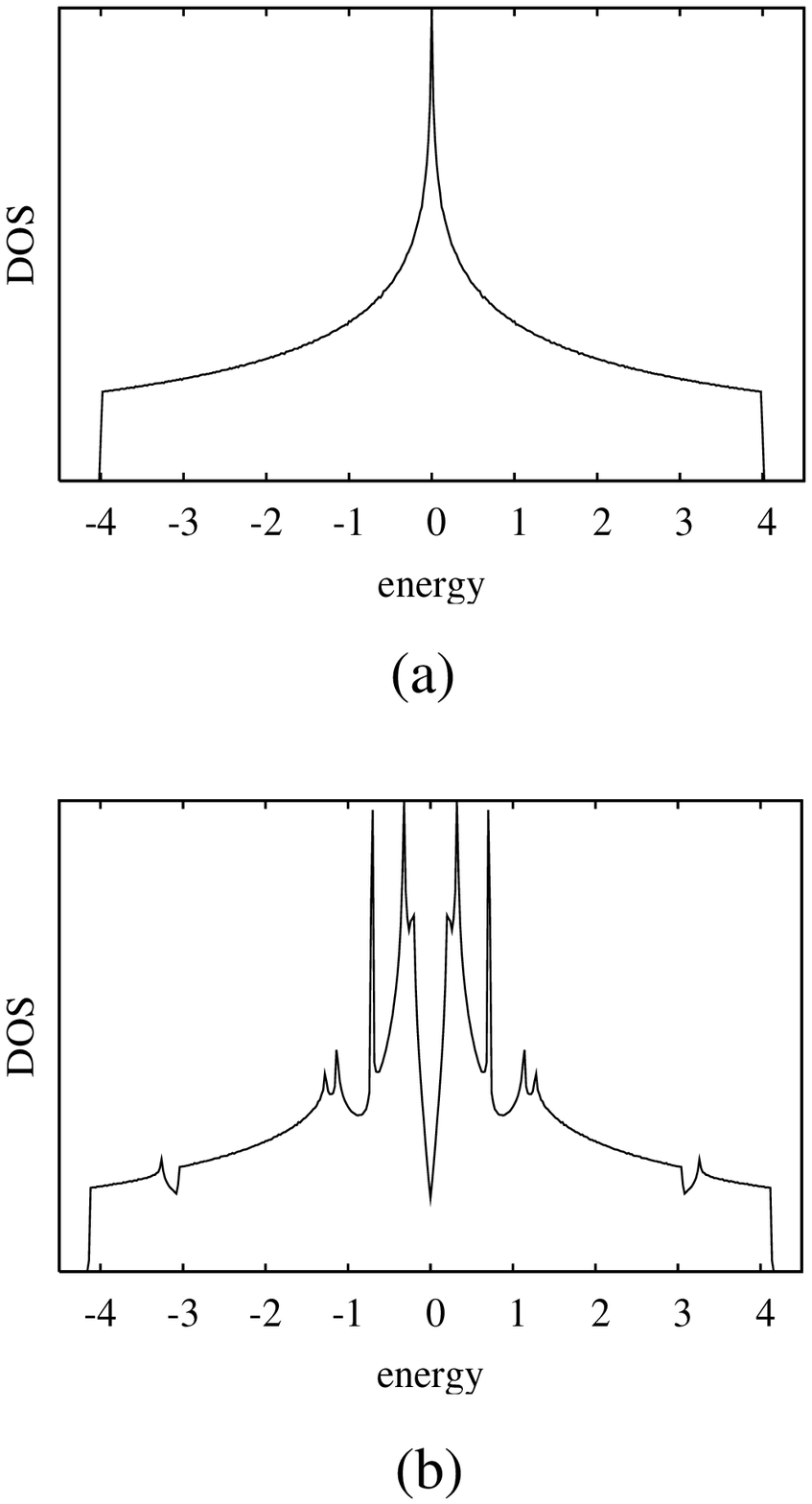}
\end{center}
\caption{(a) Density of states (DOS) of the square lattice.  
(b) DOS of the most stable state in the three sublattice states 
with the $6 \times 6$ supercell in the thermodynamic limit.
}
\label{fig:dos}
\end{figure}

Let us consider which state, the AF chain stacked state 
or the three sublattice state, is energetically stable. 
Numerical calculation 
shows that the smallest energy state in the three sublattice states
is more stable than the AF chain stacked states. 
The energy per sites for the AF chain stacked states and 
that for the three sublattice states are 
-0.811$t$ and -0.813$t$, respectively. 
The DOS of the AF chain stacked states and 
that of the most stable three sublattice state 
are shown in Fig. \ref{fig:dos}(a) and Fig. \ref{fig:dos}(b). 
In Fig. \ref{fig:dos}(b), we find several dips or imperfect gaps. 
It is considered that such dips are ascribed to the periodic spin structure, 
that is, the band gap is partially opened 
by the breaking of the translational symmetry. 
On the other hand, any gaps have not been observed in the case of 
the AF chain stacked states in Fig. \ref{fig:dos}(a). 
Thus, the three sublattice state is expected to be stabilized 
by the gap opening due to the appearance of a certain periodicity.

As mentioned previously, in the AF Ising model, 
the ground state is determined by maximizing $N_{\rm AF}$, 
while it is quite plausible but still not certain 
whether the ground state of the AF-DE model 
has the maximum number of  $N_{\rm AF}$. 
In order to search the ground state in the AF-DE model 
on the triangular lattice, 
energies of all possible spin configurations for small systems 
with $3 \times 3$ and $4 \times 4$ supercells in the thermodynamic limit 
have been calculated. 
As a result, it is found that 
any spin configurations except for the AF chain stacked states and 
the three sublattice states cannot be stabilized. 
It seems that 
the ground state has the maximum number of 
$N_{\rm AF}$ also in the AF-DE model.

Throughout this paper, two types of the ground states 
in the AF Ising model have been considered, i.e., 
the AF-chain stacked states and the three sublattice states, 
as the possible candidates for the ground state of the AF-DE model. 
As far as we have searched the lowest energy state 
among such two kinds of states for small system sizes 
when $t'=0$ and $\mu=0$, 
it is found that 
the ground state is selected from the three sublattice states. 
Such a ground state might also be affected by $\mu$. 
The DOS shown in Fig. \ref{fig:dos}(b) has the largest dip around $\mu =0$, 
which may stabilize the ground state shown in Fig. \ref{fig:GSAFDE}. 
Changing $\mu$ in the AF-DE system, 
other periodic patterns in the three sublattice states 
can emerge as the ground state by opening the gap around $\mu$. 
On the other hand, it is considered that 
the stability of the AF chain stacked states does not depend on $\mu$. 
Thus, at least two possibilities are considered: 
(1) The ground states with commensurate periodic structures can be realized 
only at a certain $\mu$, and otherwise, 
the AF chain stacked states appear in the ground states 
with degeneracy. 
(2) At each $\mu$, ground states with possible incommensurate periodicities 
corresponding to $\mu$ are selected from the three sublattice states, 
and thus, the ground state does not have the degeneracy in all $\mu$ range. 
More detailed analyses are needed to clarify this point.


In conclusion, as far as the AF-chain stacked states and 
the three sublattice states are considered, 
the lowest energy state for the system with $6 \times 6$ supercells 
can be selected uniquely from the latter states. 
Thus, degenerate ground states in the AF Ising model 
can be lifted by the effect of motions of electorns in the AF-DE model.
In this way, according to the nature of the interaction, 
the ground state is drastically changed. 
In the frustrated electron system, 
the kinetics of electrons plays an important role 
to lift the degeneracy due to the frustration and 
possibly leads to an ordered state.

\end{document}